# Improving the Resolution and Throughput of Achromatic Talbot Lithography

Dimitrios Kazazis[a)], Li-Ting Tseng, Yasin Ekinci

Paul Scherrer Institute, 5232 Villigen-PSI, Switzerland

[a)] Electronic mail: dimitrios.kazazis@psi.ch

High-resolution patterning of periodic structures over large areas has several applications in science and technology. One such method, based on the long-known Talbot effect observed with diffraction gratings, is achromatic Talbot lithography (ATL). This method offers many advantages over other techniques, such as high resolution, large depth of focus, high throughput, etc. Although the technique has been studied in the past, its limits have not yet been explored. Increasing the efficiency and the resolution of the method is essential and might enable many applications in science and technology. In this work, we combine this technique with spatially coherent and quasi-monochromatic light at extreme ultraviolet (EUV) wavelengths and explore new mask design schemes in order to enhance its throughput and resolution. We report on simulations of various mask designs in order to explore their efficiency. Advanced and optimized nanofabrication techniques have to be utilized to achieve high quality and efficient masks for ATL. Exposures using coherent EUV radiation from the Swiss light source (SLS) have been performed, pushing the resolution limits of the technique for dense hole or dot patterning down to 40 nm pitch. In addition, through extensive simulations, alternative mask designs with rings instead of holes are explored for the efficient patterning of hole/dot arrays. We show that these rings exhibit similar aerial images to hole arrays, while enabling higher efficiency and thereby increased throughput for ATL exposures. The mask designs with rings show that they are less prone to problems associated with pattern collapse during the nanofabrication process and therefore are promising for achieving higher resolution.

## I. INTRODUCTION

Patterning at the nanoscale has been of substantial interest in the recent decades especially for applications in electronic devices and integrated circuits, as CMOS downscaling dictated by Moore's law has been pushing feature sizes deep into the nanoscale.[1] At the same time there is strong interest in the scientific community for patterning of periodic structures over relatively large areas, and with high resolution, while keeping the exposure efficiency or aerial throughput high, as for example in graphene antidote lattices,[2] photonic crystals,[3] magnonics,[4] phononics,[5] plasmonics,[6] directed self-assembly applications,[7] photovoltaics[8] etc. In addition, high-resolution periodic patterning at extreme ultraviolet (EUV) wavelengths, has recently been extremely important to both industry and academia for the development and testing of EUV photoresist materials,[9] as access to EUV scanners is limited and very expensive.

In most research projects and applications requiring periodic patterning, direct write methods have been utilized such as electron beam lithography (EBL), focused ion beam (FIB) lithography, scanning probe lithography (SPL), and scanning tunneling microscope (STM) lithography. Unfortunately, all these direct write lithography techniques suffer from the same issue. There is a power-law relationship between areal throughput ($A_t$) and achievable resolution ($R$), known as Tennant's law:[10]

$$A_t = k_T R^5 , \quad (1)$$

where $k_T$ is a constant. What this law simply dictates is that reducing the resolution by a factor of 2, causes a decrease in throughput by a factor of 32, making the aforementioned direct write techniques inefficient for large-area and high-resolution patterning. In addition to the low throughput of the serial patterning methods, they have further disadvantages and limitations. For example, EBL, one of the most widely-used patterning methods, suffers from proximity effects, due to electron backscattering, charging, depending on the substrate conductivity, and a very limited depth of focus. One alternative, non-direct write technique with high throughput is nanoimprint lithography (NIL),[11] which is a one-to-one replication technique. However NIL still relies on the fabrication of a master with the same resolution as the replicas and therefore itself benefits from other high-resolution and large-area lithographic methods.

There is, therefore, a high demand for photon-based and parallel lithography techniques that are capable of producing high-resolution periodic patterns with high throughput. In this context, EUV interference lithography (EUV-IL) has emerged as one such technique and several tools based on synchrotron,[12,13] laser,[14,15] and plasma sources[16] have been set up, demonstrating high-resolution patterning down to 6 nm (dense line/space patterns).[17,18] Whether spatially coherent synchrotron light is used with multiple transmission diffraction gratings[17] or temporally coherent light in a Lloyd mirror interferometer setup,[14] stitching of multiple fields is not straightforward, due to the exposure of the photoresist by the non-interfering diffraction orders. Moreover, with these techniques a significant part of the incident beam is absorbed by photon stop layers or diffracted into unused (non-interfering) beams, thus reducing the efficiency of the method.[19]

An alternative parallel patterning method for producing periodic structures that is related to interference lithography, is the achromatic Talbot lithography (ATL), also known as achromatic spatial frequency multiplication (ASFM).[20] This technique is based on the Talbot effect, a self-imaging property of diffraction gratings first observed by H. F. Talbot in 1836[21] and analytically explained later by L. Rayleigh.[22] When coherent monochromatic light is incident upon a diffraction grating, self-images of the grating appear at well-defined distances from the grating, known as Talbot distances:

$$Z_{T,n} = \frac{\lambda}{1-\sqrt{1-\lambda^2/p^2}} n , \quad (2)$$

where $n$ is an integer, $p$ is the grating period, and $\lambda$ is the light wavelength. The above equation simplifies to:

$$Z_{T,n} = \frac{2p^2}{\lambda} n , \quad (3)$$

when $\lambda \ll p$. At $z=Z_T/2$ a self-image is also produced but laterally shifted by $p/2$ with respect to the grating. These images can be understood as the result of a phase shift of $\pi$ of the odd-numbered diffraction orders with respect to the zero and even-numbered orders.[23] At $Z_T/4$ a self-image of the grating is also produced but with a period of $p/2$ and so on for rational multiples of $Z_T$, creating a fractal pattern of images known as the Talbot carpet.[23] Although the Talbot effect has been successfully used to lithographically produce submicron resolution patterns,[24] its limited depth-of-focus requires very precise positioning of the sample with respect to the grating and makes it very challenging for high-resolution nanopatterning. A technique that is based on the Talbot effect, but circumvents this issue, is displacement Talbot lithography (DTL).[25] In this technique, the mask is illuminated with a monochromatic beam and the distance between the mask and the substrate changes during the exposure by a Talbot period. This results in an aerial image that doesn't require precise positioning at the Talbot distances, but only a displacement of the mask or substrate by a Talbot period independently of the start and end positions. If the illumination is not monochromatic, as it will be shown in the next section, under certain conditions, a stationary image with large depth-of-focus can be obtained, enabling the ATL or ASFM technique for lithographic patterning at the nanoscale.

The ATL technique has several advantages that enable it as a powerful lithographic technique. First, the resulting stationary image is characterized by frequency multiplication. For example, in the case of a one dimensional grating, the period of the stationary image is half of the grating period. Examples of frequency multiplication will be shown in the simulations in the subsequent section. Moreover, as the stationary image is independent of the wafer-mask distance, we can achieve very large depth-of-focus. This enables patterning over large areas, without worrying about wafer thickness inhomogeneities, and on three dimensional surfaces, where other methods like EBL or standard EUV-IL are limited. In addition, as all the diffracted orders contribute to the aerial image, ATL exposures are highly efficient, unlike multiple grating EUV-IL, where part of the beam is absorbed by photon stop layers or diffracted into non-interfering beams. Furthermore, as only one transmission grating is necessary, there are no unwanted diffraction orders exposing the photoresist around the useful pattern, as for example in the case of EUV-IL, where the zero orders and the first or higher non-interfering orders are also printed on the resist. This makes large area patterning possible by step and repeat. It should be noted that it is possible to stitch multiple fields also in the case of multiple beam interference lithography, by utilizing an order-sorting aperture.[26] Nevertheless, the single-exposure patterned area in the multiple beam case is much smaller than in the ATL case, as the mask needs to accommodate all of the gratings. Another major advantage of ATL is its self-healing property.[27] Small local defects on the transmission grating are not printed, since the aerial image is formed by the light diffracted from a large area of the mask. Last, as explained in detail by Guérineau et al.[28] for gratings with a rectangular cross section, a triangular intensity profile is obtained. This leads to a constant aerial image contrast, i.e. normalized image log slope, enabling the patterning of small features by controlling the exposure dose.[29] Thus, one can tune the duty cycle, i.e. linewidth/period and make it considerably smaller than in the case of multiple beam interference.

Given all the aforementioned advantages of ATL, its further improvement in terms of resolution and throughput is of scientific and technological interest and we believe that its potential has not been fully explored. In this paper, we demonstrate how the aerial image in ATL is formed and how to improve its resolution and throughput. We start with the theory of ATL and simulations of aerial images of different mask designs. In the subsequent section, we present the experimental results on the nanofabrication of ATL transmission masks and ATL exposures at the SLS. In the final section, we summarize our major results and discuss the future perspectives of our work on ATL for improved resolution and efficiency.

## II. THEORY AND SIMULATIONS

In this section, we discuss the Talbot effect in the quasi-monochromatic regime and describe our method to simulate the aerial images of given mask configurations. As already discussed in the previous section, Talbot effect is the interference of the diffracted orders of a grating, which leads to the formation of self-images of it at specific locations given by Eq. (3). In the case of non-monochromatic light, the self-images at the Talbot planes described by Eq. (3), are no longer well defined, but are smeared out between the two values:

$$Z_{T,n} = \frac{2p^2}{\lambda \pm \frac{\Delta\lambda}{2}} n, \qquad (4)$$

as dictated by the light bandwidth $\Delta\lambda$. With increasing $n$, when the upper part of the $n^{th}$ Talbot plane image (corresponding to $\lambda-\Delta\lambda/2$):

$$Z_{T,n}^{max} = \frac{2p^2}{\lambda - \frac{\Delta\lambda}{2}} n \qquad (5)$$

reaches the bottom part of the $n+1$ Talbot plane image (corresponding to $\lambda+\Delta\lambda/2$):

$$Z_{T,n+1}^{min} = \frac{2p^2}{\lambda + \frac{\Delta\lambda}{2}} (n+1), \qquad (6)$$

then the smeared out images merge together, resulting in a stationary image. A similar analysis holds for the $p/2$-shifted self-images in between the Talbot planes, which also smear out and merge, contributing to the stationary image and resulting in the frequency multiplication. This is shown schematically in Fig. 1. By equating Eqs. (5) and (6) it is easily shown that the stationary image is obtained when:

$$n \geq \frac{\lambda - \frac{\Delta\lambda}{2}}{\Delta\lambda}. \qquad (7)$$

This corresponds to a minimum distance from the grating, after which the intensity is stationary, called achromatic Talbot distance $Z_A$. Plugging Eq. (7) into Eq. (5) we obtain:

$$Z_A = \frac{2p^2}{\Delta\lambda}. \quad (8)$$

In general, the maximum distance $Z_{max}$ up to which the stationary interference pattern is obtained is limited by the spatial coherence of the incident illumination.[20] For a given transverse coherence length $l_c$ of the beam:

$$Z_{max} = \frac{l_c p}{2\lambda}. \quad (9)$$

This maximum distance is particularly important for partially coherent sources, as for example for plasma-based radiation sources, where $l_c$ is only a few micrometers long.[30] In the case of synchrotron sources, spatial coherence is large, on the order of several millimeters and does not pose a limitation to the extent of the stationary image. In this case it is only limited by purely geometrical factors, as ATL is a consequence of interference of diffracted orders of a single grating. The overlap of these orders and thereby the patterned area decreases with increasing distance from the grating and vanishes at a distance:[31]

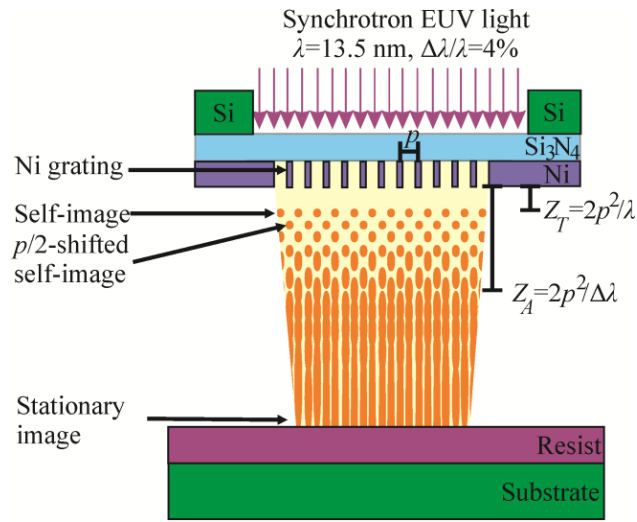

**FIG. 1.** Schematic view of the achromatic Talbot lithography setup. The self-images smear out due to the bandwidth $\Delta\lambda$ and eventually merge after the achromatic distance $Z_A$, creating a stationary image.

$$Z_{max} = \frac{Wp}{2\lambda}, \quad (10)$$

where $W$ is the size of the diffraction grating.

Although it is semi-intuitive to understand how an aerial image is formed and what it would look like in the case of ATL for simple mask designs like linear gratings, it is certainly not always the case, as it will be explained in this section. Here we will present a simple but accurate method to simulate the aerial images of any arbitrary mask design. These simulations provide a helpful guide for designing efficient ATL masks with high resolution as well as considering feasibility of their fabrication.

For this purpose we calculate the propagation of the diffracted light through a transmission diffraction grating. We assume that the incident light is described by a plane wave:

$$U(x, y, z) = A(x, y, z) e^{j\phi(x,y,z)}, \quad (11)$$

where $A$ is the amplitude and $\varphi$ the phase at position $(x,y,z)$. We assume rectangular profiles for the gratings and wave propagation in the Fresnel approximation. In this approximation, for a field $U_1(x,y)$ at the source, the field $U_2(x,y)$ at the observation plane is given by:

$$U_2(x, y) = \Im^{-1}\{\Im\{U_1(x, y)\} H(f_X, f_Y)\}, \quad (12)$$

where $\Im$ and $\Im^{-1}$ are the Fourier and the inverse Fourier transform respectively and $H$ is the transfer function:

$$H(f_X, f_Y) = e^{jkz} e^{-j\pi\lambda z(f_X^2 + f_Y^2)}, \quad (13)$$

where $k=2\pi/\lambda$ is the wavenumber, $z$ is the distance between source and observation plane, and $f_X$ and $f_Y$ are the Fourier domain frequencies associated with $x$ and $y$ axes respectively. The aerial image, which is recorded in the photoresist during a lithographic process, is the irradiance or intensity of the light, which is:

$$I_2(x, y) = U_2(x, y)U_2(x, y)^* = |U_2(x, y)|^2. \quad (14)$$

In this analysis, we have assumed monochromatic light at wavelength $\lambda$. In order to account for the quasi-monochromatic light with bandwidth $\Delta\lambda$, which is essential for ATL, we consider Gaussian power spectral density:

$$\hat{S}(\nu) = \frac{1}{\sqrt{\pi}b} e^{\left[-\frac{(\nu-\nu_0)^2}{b^2}\right]}, \quad (15)$$

with:

$$b = \frac{\Delta\nu}{2\sqrt{\ln 2}}, \quad (16)$$

where $\Delta\nu$ is the full width at half-maximum of the Gaussian power spectral density. The total intensity is calculated as the integral over all the spectral components, which can be approximated by the sum of all the discrete components for computational simplicity:

$$I(x, y) = \int_{-\infty}^{+\infty} \hat{S}(\nu)I(x, y;\nu)d\nu \approx \sum_{n=1}^{N} \hat{S}(\nu)I(x, y;\nu_n)\delta\nu. \quad (17)$$

In the present study, we have assumed 3600 spectral components. Numerical computations were performed in the MATLAB numerical computing environment (MathWorks, Inc.). The input to the implemented simulation program is the unit cell of a periodic mask pattern in the format of a grayscale image or matrix, which can be created computationally or in any CAD software. The fast Fourier transform (FFT) algorithm that is used in Eq. (12), takes care of the periodic extension of this unit cell to create an infinitely large diffraction mask. The output of the algorithm is the aerial image, the intensity of the light at the observation plane far enough from the mask, at distances greater than the achromatic distance $Z_A$ of Eq. (8), typically at $2Z_A$. The algorithm produces the unit cell of the periodic aerial image. We note that in this study, we did not consider the optical properties of the mask or diffraction effects within the mask, but assumed that the mask is an absorber, in which case the amplitude of the optical field behind the mask is the binary mask layout.

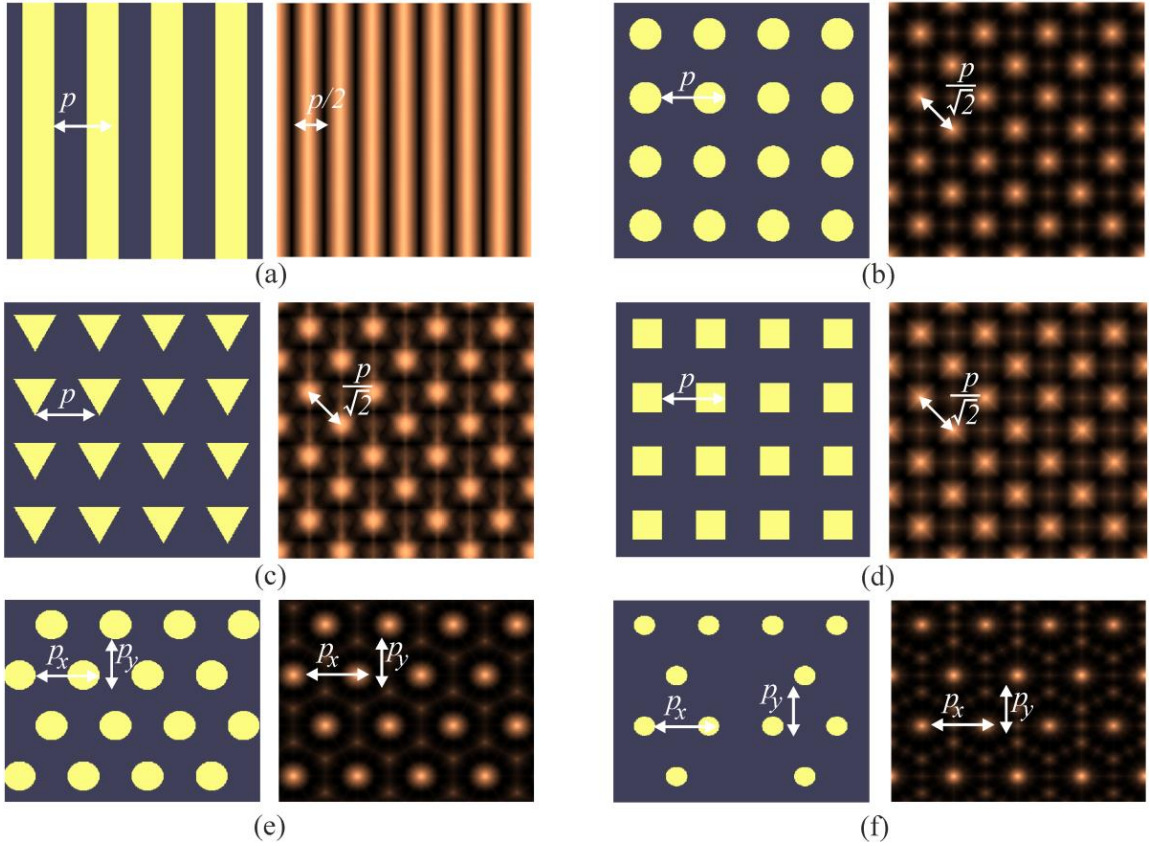

**FIG. 2.** Typical examples of mask designs (left images) and simulated aerial images (right), at mask-observation plane distances beyond the achromatic distance $Z_A$. (a) A linear grating produces a linear aerial image with half the period. Masks with square arrays of (b) holes, (c) triangles, and (d) squares result in aerial images with 45° rotated intensity peaks with pitch divided by $\sqrt{2}$. Masks with hexagonal arrays of holes as in (e) and (f) produce hexagonal aerial images with the same pitches.

Examples of simulations are shown in Fig. 2, where the resulting aerial images (on the right) are shown together with the input transmission masks (on the left). Since the input to the code is a mask layout, more versatile patterns can also be simulated, only limited by the imagination of the mask designer. The importance of the simulator is paramount, as it is not always straightforward to predict the aerial image for a specific transmission mask.

In the following, we will concentrate on the simulation of masks that result in arrays of holes or dots like for example the one shown in Fig. 2 (b), as there are numerous applications requiring dense dot or hole arrays over relatively large areas. The aerial image in this case is an array of dots rotated by 45$^o$ with a pitch divided by √2. Around these main peaks, at a distance of pitch/2 along all axes we see some secondary, satellite peaks of lower intensity. These peaks are due to higher diffraction orders of the grating mask. Since the lithographic process is the conversion of the grayscale aerial image into a binary image, small peaks and low-intensity features are not printed at low doses.

There are two directions to follow, both of which are significant. The first direction is to increase the resolution, i.e. produce smaller and denser dots. The second direction is to increase the mask efficiency. Any light source has a limited flux and it is therefore essential that the masks can pattern the photoresists with the required dose in reasonable times. The significance of this second direction becomes evident in the simulations presented in Fig. 3, where the effect of the dot size on the intensity of the aerial image is presented. Simulations of aerial images for masks with hole arrays with hole diameters ranging between 50 and 300 nm and a relaxed pitch of 1 μm are shown. The mask and aerial images for a 300 and 50 nm hole are shown in Fig. 3 (a) and (b). In

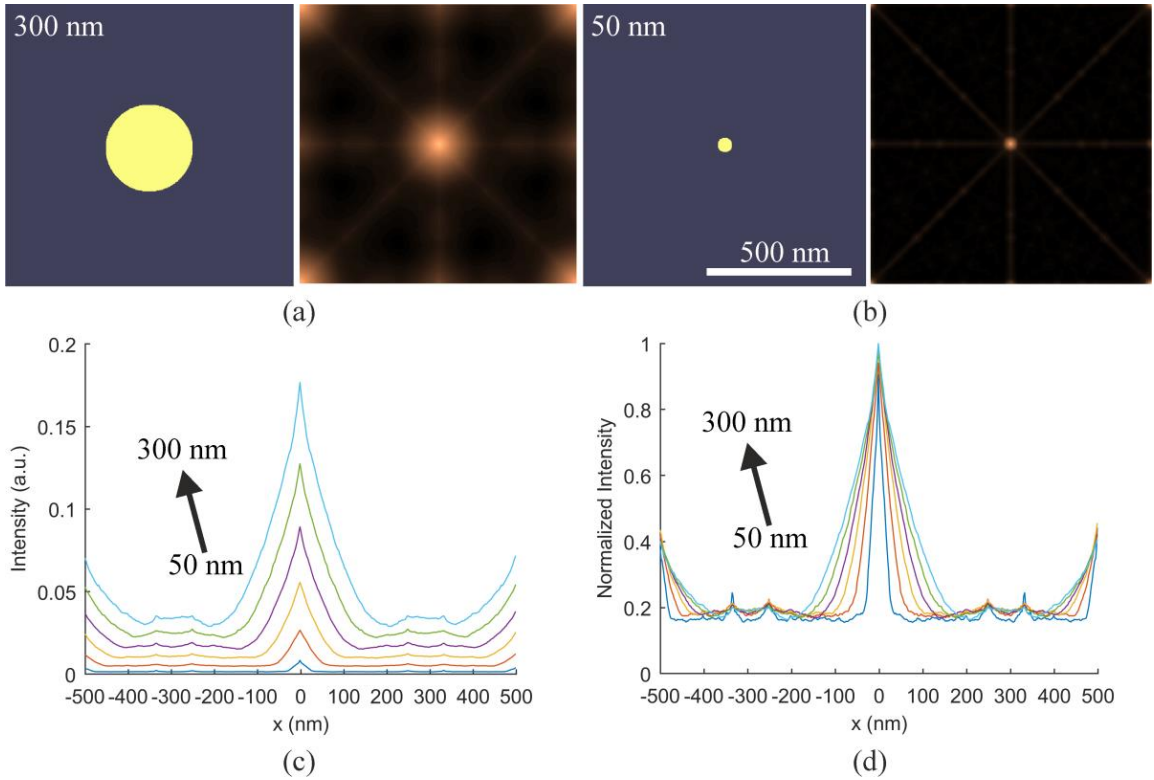

**FIG. 3.** Simulations of the aerial images produced by an array of (a) 300 nm diameter holes and (b) 50 nm diameter holes. (c) Unnormalized intensities along the horizontal line passing through the holes for several hole sizes (50, 100, 200, 300, 400, and 500 nm) showing the decrease in the intensity for the small holes. (d) The normalized curves of (c) showing the effect of the hole size on the width of the intensity peak. The simulations assume a pitch of 1 μm and a mask-observation plane distance of twice the achromatic distance: $z=2Z_A=7.41$ mm. The shown scale bar applies to all images in (a) and (b).

Fig. 3 (c) the unnormalized intensities along the horizontal line passing through the center of the intensity peaks are plotted. It is clear that by decreasing the hole size, to achieve smaller dots on the aerial image, we also decrease the obtained intensity and therefore the efficiency of the mask. Figure 3 (d) shows the same curves with the intensity normalized to the maximum of each peak. Here, the effect of the mask hole size on the width of the intensity peak is illustrated, with the smaller holes resulting in narrower and sharper peaks. The above observations lead to the conclusion that even though reducing the hole size on the mask leads to smaller feature sizes, this also leads to a dramatic reduction in the obtained intensity and thus in a reduced efficiency of the exposure and low lithographic throughput.

In the quest for more efficient and high-resolution mask designs, we have explored various geometries and their resulting aerial images, which is enabled by the simulator that can quickly simulate aerial images of various unconventional mask designs. For example, we discovered that the aerial image of a circular ring is, to our surprise, a dot. This can be seen in Fig. 4 (a) and (b) where the mask and corresponding aerial images of circular rings with different widths are shown. The first series of simulations that we performed were to study the effect of the ring width. We simulated rings of a fixed diameter of 500 nm, at a pitch of 1μm, with varying ring width from 25 up to 200 nm. Figure 4 (c) shows the unnormalized intensity profiles along the horizontal line passing through the center of the rings. It is clear that increasing the ring diameter results in higher intensities. Nevertheless, if we look at the normalized curves in Fig. 4 (d), it is clear that increasing the ring width increases the width of the intensity peak and therefore degrades the mask resolution.

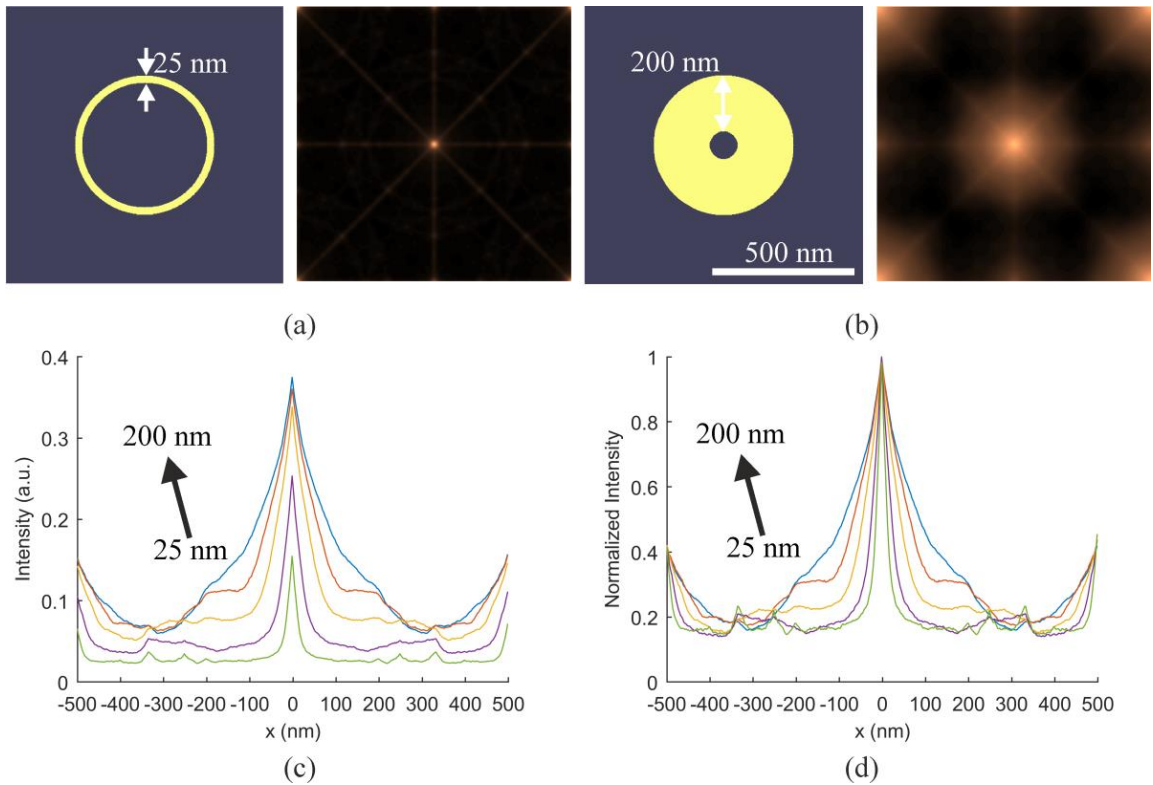

**FIG. 4.** Simulations of the aerial images produced by an array of 500 nm diameter circular rings with a width of (a) 25 nm and (b) 200 nm. (c) Unnormalized intensities along the horizontal line passing through the center of 500 nm diameter rings with several widths (25, 50, 100, 150, 200 nm) showing the increase in the intensity for the rings with a larger width. (d) The normalized curves of (c) showing the effect of the ring width on the width of the intensity peak. The simulations assume a pitch of 1 μm and a mask-observation plane distance of twice the achromatic distance: $z=2Z_A=7.41$ mm. The shown scale bar applies to all images in (a) and (b).

Since varying the width of the ring does not result in the desired intensity profiles (narrow peaks with increased intensity), we decided to keep the width of the ring fixed at a relatively small value (25 nm) and study the effect of varying the ring diameter. Increasing the ring diameter has almost no effect on the intensity peak width and even with the largest diameter rings narrow peaks can be obtained, as shown in Figs. 5 (a) and (b) where a large 800 nm diameter ring and a smaller 100 nm diameter ring are respectively simulated. The aerial images are almost identical and very similar to the aerial image of the 50 nm hole depicted in Fig. 3 (b). A comparison of the unnormalized intensities of the different rings shown in Fig. 5 (c) makes it clear that there is a large increase in the intensity, with increasing ring diameter. The largest ring gives the highest intensity and the 50 nm hole the lowest. The 50 nm hole intensity is also included in the graph (lowest lying curve) for comparisons. What is even more interesting is to look at the normalized intensities of the rings, shown in Fig. 5 (d). It is clear that it is the width of the ring that defines the intensity peak width and not the diameter of the ring. The widths

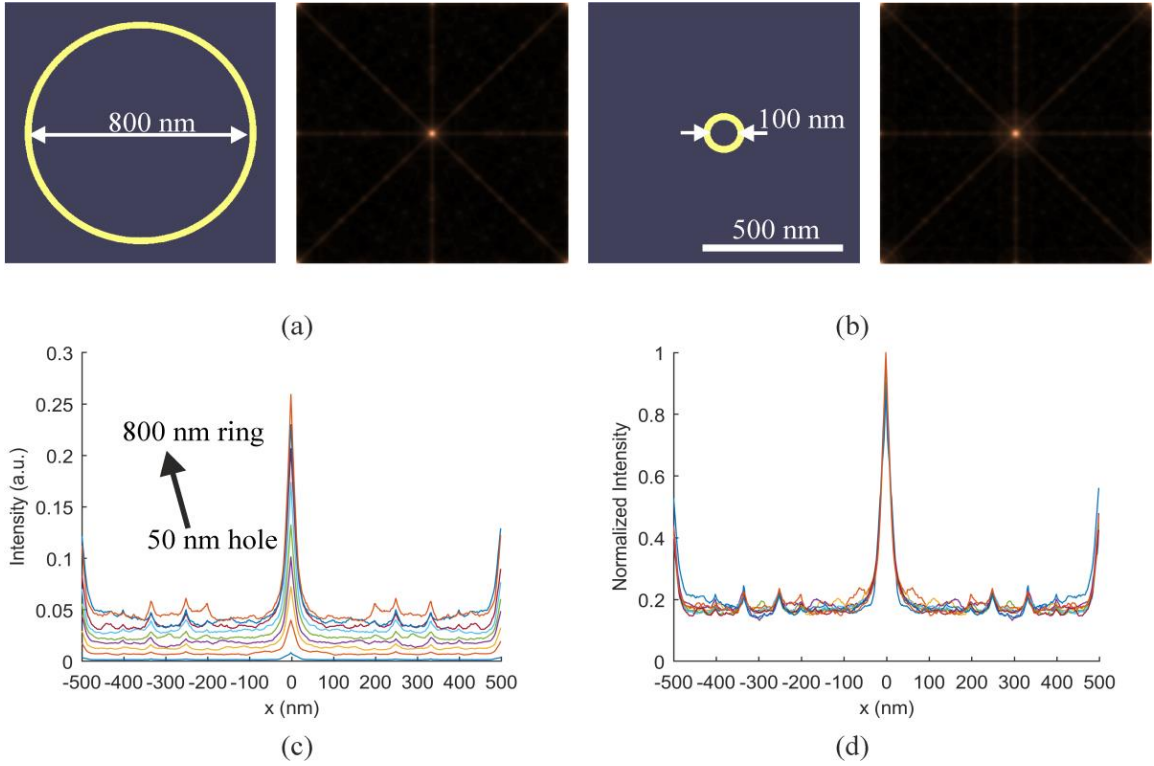

**FIG. 5.** Simulations of the aerial images of an array of 25 nm wide circular rings with a diameter of (a) 800 nm and (b) 100 nm. (c) Unnormalized intensities along the horizontal line passing through the center of several diameter rings (100-800 nm in steps of 100 nm) showing the increase in the intensity for the rings with a larger diameter. The intensity of the 50 nm hole (lowest intensity) is also included in the plot for comparison. (d) The normalized curves of (c) show that all the rings produce intensity peaks of the same width. For comparison, the intensity of the 50 nm hole of Fig. 3 is included in the graph. The simulations assume a pitch of 1 μm and a mask-observation plane distance of twice the achromatic distance: $z=2Z_A=7.41$ mm. The shown scale bar applies to all images in of the normalized intensities of all the rings are almost identical and almost the same as the intensity of the 50 nm hole (also shown in the graph for comparison).

To conclude this section, we have shown that it is possible to obtain sharp and narrow intensity peaks, similar to the ones obtained by masks with hole arrays, by replacing the holes with rings of larger diameter with narrow width. Whereas the circular apertures lead to a tradeoff between the resolution and efficiency, ring design substantially overcomes this limitation by providing high efficiency and resolution simultaneously, which is important for several applications. However when it comes to applications where dense patterns are required, for example with pitches below 100 nm, the ring geometries presented in this section start to become complicated and eventually impossible from a nanofabrication perspective. It is hence important on one hand to pursue the fabrication of masks with ring geometries for increased efficiency and on the other hand to further investigate, by means of simulations, even more complicated geometries involving rings for higher-density patterns.

## III. EXPERIMENTAL RESULTS AND DISCUSSION

Transmission diffraction masks like the ones shown in Fig. 1 for use in ATL exposures have been fabricated before.[32] Here we present an optimized process for the nanofabrication of masks with dense hole arrays of higher resolution. The nanofabrication scheme is outlined in Fig. 6. The transmission masks are fabricated on thin, suspended, low-stress $Si_3N_4$ membranes typically around or less than 100 nm thick and 3x3 $mm^2$ large, as shown in Fig. 6 (a). Such membranes have extensively been used in the fabrication of diffractive optics and mask gratings for EUV light, due to their reasonable transmission at the 13.5 nm wavelength. A blanket Ti/Au/Cr metal deposition follows, as in Fig. 6 (b). The Ti and Cr layers are used due to their excellent adhesion-promoting property and Au is used as the seed layer for electroplating. Subsequently, hydrogen silsesquioxane (HSQ) resist is spin-coated on the metal layer and exposed by EBL, in an optimized beam stepping strategy in order to obtain nicely circular HSQ dots. After developing in a salty developer solution (1 part Microchemicals AZ 351B to 3 parts deionized water), the masks are dried using supercritical drying so as to avoid the HSQ pillar collapse. The result is schematically shown in Fig. 6 (c). The HSQ pillars can have diameters down to 28 nm, with a 50% duty cycle (pitch of 56 nm) and a high aspect ratio of ~5 that corresponds to a height of more than 150 nm. In the next step the top Cr layer is removed by chlorine dry etching and a Ni film of more than 120 nm is grown by electrochemical deposition, using the HSQ pillars as a mold, as in Fig. 6 (d). In the final step, the HSQ pillars are removed in a buffered HF solution and the remaining Ti/Au/Cr layers under the HSQ are removed by means of chlorine dry etching (Cr) and Ar ion milling (Au). During the Au etch by ion milling (about 20 nm), the Ni layer is also attacked and etched, as Ar ion milling is not a selective process. But Ni is etched at a slower rate than Au (roughly half), so we expect less than 10 nm of the Ni mask to be etched. This is compensated for during the electroplating step, where a thicker Ni film is grown in advance. The remaining Ti underlayer is very thin (< 5nm) and fairly transparent at the 13.5 nm wavelength and therefore it poses no problem if it is not etched. The final mask with the electroplated Ni absorber and the square array of cavities is shown schematically in Fig. 6 (e).

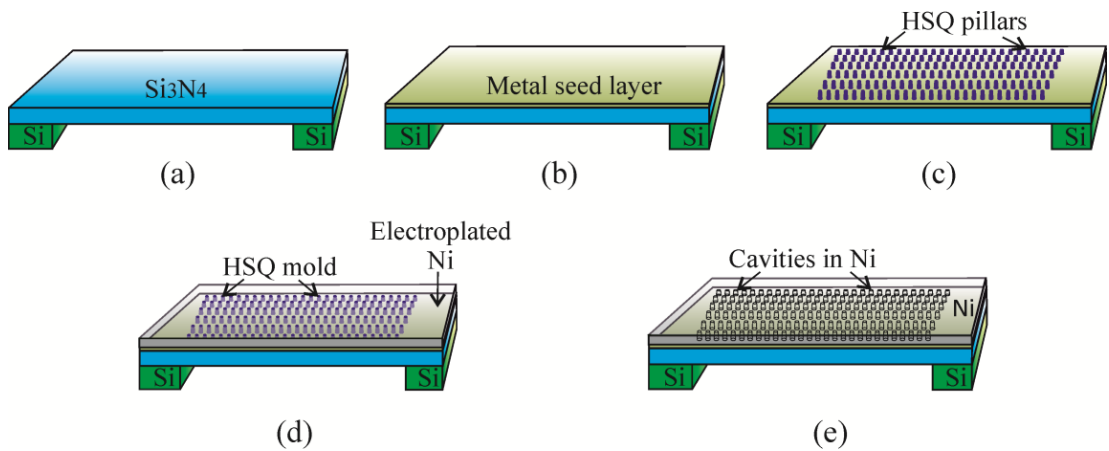

**FIG. 6.** Nanofabrication process for ATL masks. (a) A suspended low-stress $Si_3N_4$ membrane. (b) Blanket deposition of a thin Ti/Au/Cr metal layer used as a seed layer for subsequent electrochemical deposition. (c) HSQ resist pillars are formed by EBL. (d) The HSQ pillars are used as a mold for Ni electroplating. (e) The HSQ pillars are etched in a buffered HF solution, and the metal seed layer is ion milled leaving behind the required holes in the Ni absorber film.

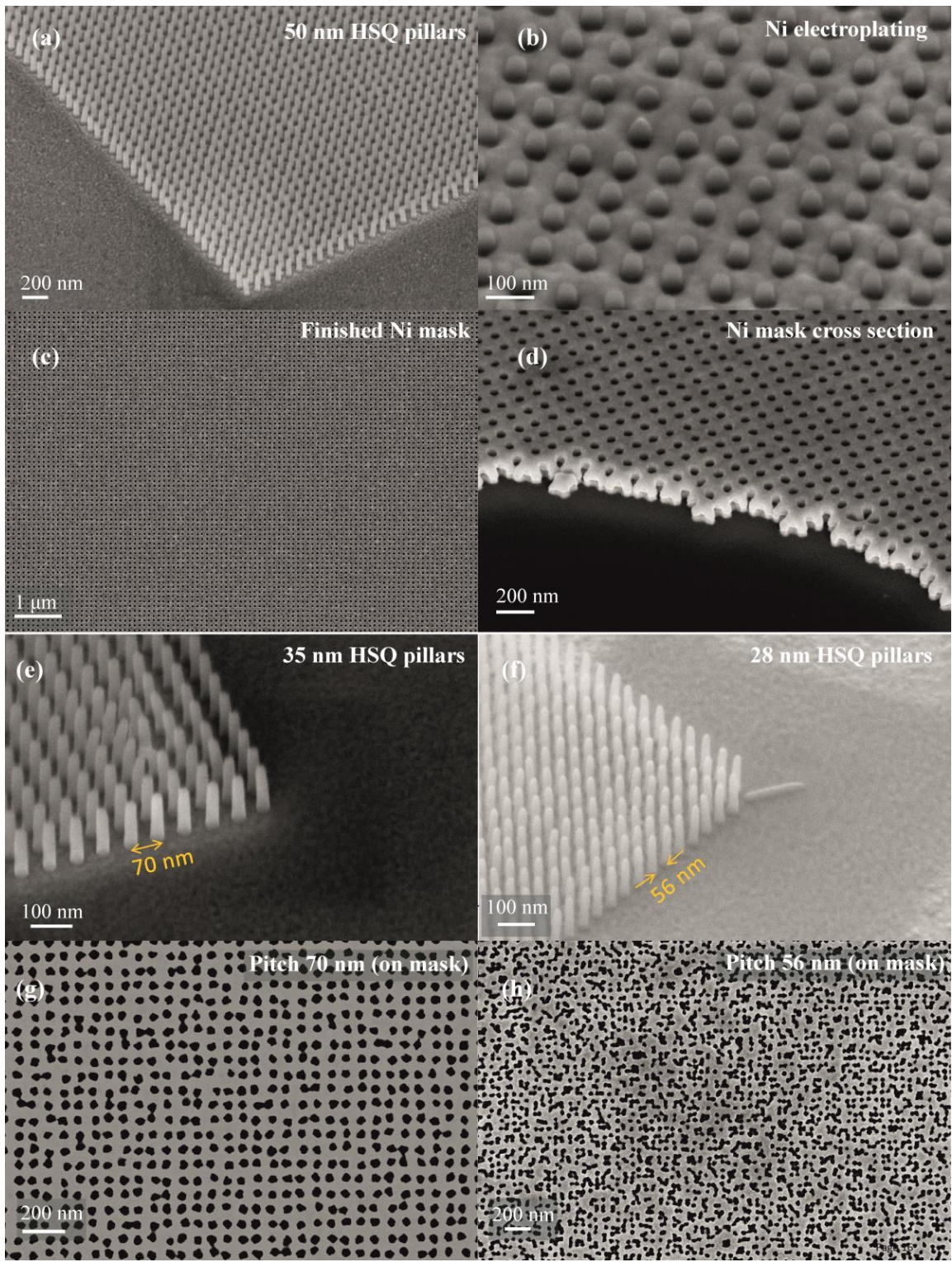

**FIG. 7.** SEM images of the masks at various nanofabrication process steps. (a) 50 nm HSQ pillars (pitch 100 nm). (b) Electroplating through the HSQ mold. (c) A finished Ni mask with 50 nm holes. (d) A cross-section of the mask in (c). (e) HSQ pillars of 36 nm (70 nm pitch). (f) HSQ pillars of 28 nm (56 nm pitch). Finished masks with 70 and 56 nm pitches are shown in (g) and (h) respectively.

Scanning electron micrographs (SEM) of fabricated masks with on-mask pitches of 100, 70, and 56 nm are shown in Fig. 7. Figures 7 (a), (e), and (f) show tilted views of the HSQ pillars after the supercritical drying step. There is no apparent pillar collapse, even in the smallest diameter ones, where the aspect ratio is more than 5. Figure 7 (b) shows the HSQ pillars of (a) sticking out of the electroplated Ni film. The finished mask for the case of the 50 nm HSQ pillars is shown in Figs. 7 (c) and (d) in a top view and a cross section respectively. Finally Figs. 7 (g) and (h) show top views of the finished masks with 36 and 28 nm pillars respectively, where the hole arrangement is not as well defined as in the case of the 50 nm pillars. This is due to the pattern collapse after the sample is put in the electroplating solution, which is worse for the smallest pillars.

There is ongoing work to face this undesirable pattern collapse when the mask pitch is pushed to smaller scales. One possible solution is to adopt an all-wet fabrication process, between the development of the HSQ after the EBL exposure and after the Ni electroplating. If the sample is not dried between the immersion into different solutions, there are little chances of collapse.[33] After the Ni film is electrodeposited, the risk of pattern collapse is significantly reduced, as the Ni film has relatively high mechanical stability. Alternative approaches to enhance the HSQ adhesion in order to prevent the pillar collapse are also investigated. Furthermore, taking into consideration the ATL simulations of the ring arrays of the previous section, we are also investigating the nanofabrication of circular rings, instead of dots, both to enhance the mechanical stability of the structures and also to increase the efficiency of the final ATL masks. An example of a first attempt to fabricate such rings is shown in Fig. 8 (a) and (b) where HSQ cylinders of 325 nm diameter and a wall width of 25 nm have been successfully defined by EBL lithography. Figures 8 (c) and (d) show that it is possible to electroplate Ni using these HSQ cylinders as a mold.

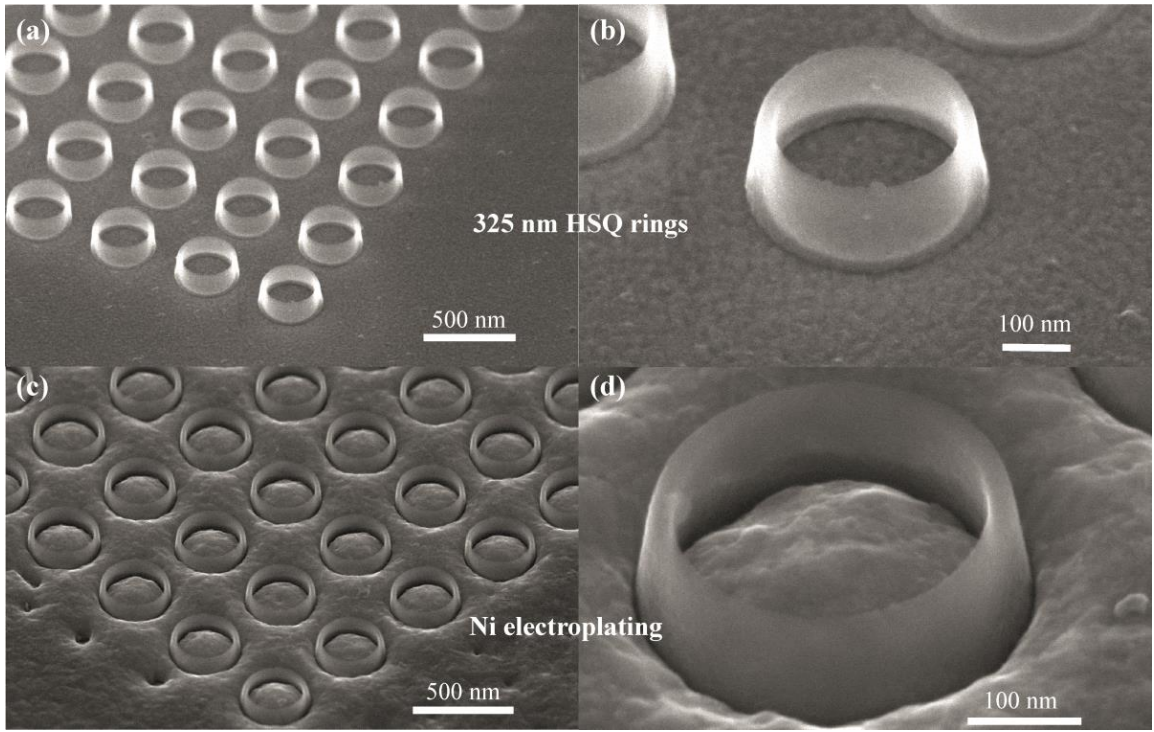

**FIG. 8.** (a) and (b) SEM images of fabricated HSQ rings of 325 nm diameter and 25 nm width. (c) and (d) Electroplated Ni film through the HSQ ring mold.

We have performed ATL exposures at the XIL-II beamline of the Swiss light source (SLS) synchrotron of the Paul Scherrer Institute (PSI) using the masks with 50 nm holes/100 nm pitch, 36 nm holes/70 nm pitch, and 28 nm holes/56 nm pitch shown in Fig. 7. The results are shown in Fig. 9. The mask-wafer distance was kept at 450 μm, which is significantly larger than the achromatic distance of the above three cases. For instance $Z_A$ is ~37 μm for 100 nm pitch mask grating. The exposure dose on mask was ~510 mJ/cm$^2$ for pitch 100 nm and pitch 70 nm and ~620 mJ/cm$^2$ for pitch 56 nm. The size of the exposed fields was 200 x 200 μm$^2$ (single exposure). This size is limited by the main-field size of the EBL tool that we used to write our masks, which is around 500 μm. For larger ATL mask sizes, EBL stitching should be employed, leading to possible non-uniformities during the ATL exposures. Although there are ways to reduce this main-field stitching during the EBL writing, by applying different exposing strategies and increase the single-exposure patterned areas, this is not a fundamental issue and within the scope of this work the smaller 200 x 200 μm$^2$ fields were sufficient.

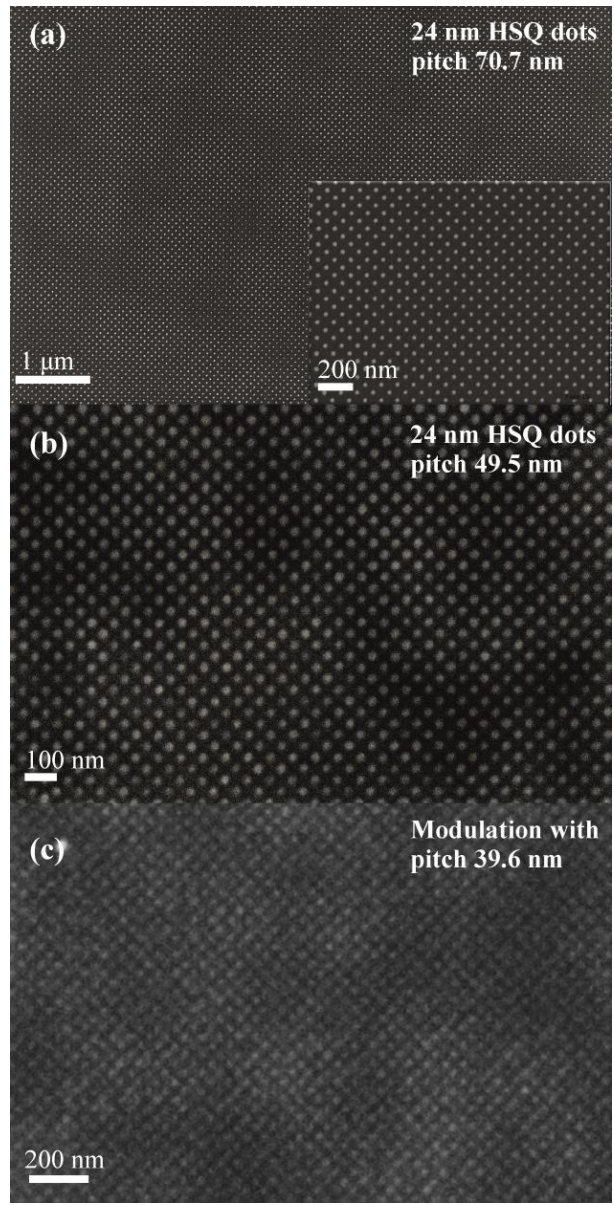

**FIG. 9.** ATL exposures using the masks of Fig. 7. The achromatic Talbot patterns are recorded in thin (~30 nm) HSQ resist. (a) SEM image of 24 nm HSQ dots with a pitch of 70.7 nm (100 nm on mask divided by √2). A zoomed-in image is shown in the inset. (b) SEM image of 24 nm HSQ dots with a pitch of 49.5 nm (70 nm on mask divided by √2). (c) A modulation with a pitch of 39.6 nm (56 nm divided by √2) is recorded on the HSQ film.

For the more relaxed mask with a pitch of 100 nm and hole sizes of 50 nm, we obtain uniform 25 nm HSQ dots with a pitch of $100\,\text{nm}/\sqrt{2} = 70.7\,\text{nm}$, as expected from the simulations of Fig. 2 (b). The results are shown in Fig. 9 (a). For the intermediate mask of 70 nm pitch and 36 nm holes, we obtained 24 nm dots with a pitch of 49.5 nm. The obtained dots in this case, shown in Fig. 9 (b) are not uniform, resulting from the nonuniformity of the mask in Fig. 7 (g) due to partial pillar collapse. For the case of the higher resolution mask of Fig. 7 (h), even though there is very little regularity in the mask pattern, we still obtain a modulation recorded in the HSQ resist, as shown in Fig. 9 (c). This is a manifestation of the aforementioned self-healing property of the ATL method,[27] where local defects on the mask do not show on the exposed photoresist, but rather add to the background of the exposure.

It is obvious from these exposures that the aerial images strongly depend on the quality of the masks. The nanofabrication of ATL transmission masks is relatively challenging. Even though it consists of a single EBL step for the definition of the mask pattern, it still involves several nanofabrication techniques like, supercritical drying, electrochemical deposition, metal depositions, reactive ion etches, Ar milling, wet etches etc. Pushing the resolution of such masks is a difficult process that relies in overcoming several issues such as the pillar collapse. Replacing the Ni with a more absorbing material would be a potential solution, as this would allow the fabrication of pillars with lower aspect ratios, and therefore less prone to collapse. As discussed above an all-wet process could also help to solve the pillar collapse issue. It should be stressed that the pillars do not collapse after the development, as supercritical drying is employed, but after the wetting of the masks during the immersion in the Ni electroplating solution.

Another method to mitigate the pillar instability is to employ more versatile structures like, for example the rings for which the simulations show impressive results and which are mechanically more stable than the relevant HSQ pillars.

## IV. CONCLUSIONS AND OUTLOOK

In summary, achromatic Talbot lithography is a very powerful lithography method, capable of producing high-resolution, periodic patterns over large areas. It is particularly useful in the case where stitching of several fields is required to pattern larger areas. Although this method has been explored for several years now, its limits have certainly not been reached yet and there are several intriguing aspects of it that should be investigated. In order to improve this method towards denser and smaller feature sizes, in this work, we have pushed the resolution and the pitch beyond what has been reported so far, by fabricating good quality and high-resolution ATL masks. The other important issue that needs attention is the mask efficiency and thereby the throughput. Patterning at reasonable times is very important and engineering ways to improve the mask efficiencies is essential. For this purpose, we have utilized easy to use but accurate modeling, which can predict the aerial image of a given mask configuration. By simulating several mask designs we have found out that ring geometries can produce very narrow intensity peaks, while not compromising the absolute intensity of the light on the photoresist. This observation opens the way to innovative mask designs that have not been explored yet. These original designs that we intend to investigate in the future have the potential to enable many applications in science and technology that require periodic nanopatterning over large areas with high throughput.

## ACKNOWLEDGMENTS

We would like to acknowledge Dr. V. Guzenko for valuable insight regarding EBL lithography and Mrs. M. Vockenhuber for her valuable assistance during the EUV exposures at the XIL-II beamline of the SLS at PSI. This project has received funding from the EU-H2020 Research and Innovation program under Grant Agreement No. 654360 NFFA-Europe.